\documentclass[11pt,twoside]{article}
\usepackage{asp2004}
\usepackage{epsf}
\usepackage{psfig}
\usepackage{lscape}
\usepackage{natbib}
\bibpunct{(}{)}{;}{a}{}{,}

\begin{document}

\title{NGC\,6231: X-ray properties of the early-type star population}
\author{H. Sana$^1$, Y. Naz\'e$^1$, E. Gosset$^1$, G. Rauw$^1$, H. Sung$^2$, J.-M. Vreux$^1$}
\affil{1. Institut d'Astrophysique et de G\'eophysique, Universit\'e de Li\`ege, All\'ee du 6 ao\^ut 17, B\^at.\ B5c, B-4000 Li\`ege, Belgium. \\
2. Department of Astronomy and Space Science, Sejong University, Kunja-dong 98, Kwangjin-gu, Seoul 143-747, Korea. \\
E-mail contact: sana@astro.ulg.ac.be }

\begin{abstract}
 Based on a deep XMM-Newton observation of the young open cluster NGC\,6231, we derive the main X-ray properties of its early-type star population. Among the 610 X-ray sources detected in the field, 42 are associated with early-type stars. We investigate their $L_\mathrm{X} / L_\mathrm{bol}$ relationship and we confirm the clear dichotomy between O- and B-type stars. The cut-off line between the two behaviours occurs at $L_\mathrm{bol} \sim 10^{38}$\,erg\,s$^{-1}$ as previously proposed by \citet{BSC97}. The distinction between single and binary stars is not clear cut, except for the colliding wind system HD\,152248. The X-ray detected B-stars in NGC\,6231 appear to be more luminous than predicted from the \citeauthor{BSC97}\ relation. Though this suggests a bimodal distribution of the B-star X-ray emission, we caution however that these results might be biased by detection limits. Finally we investigate the X-ray variability of the detected sources and we find that about 40\% of the X-ray emitters in the field of view present consistent signs of variability in the EPIC instruments. This fraction is much larger than previously thought. About one third of the early-type stars population, either singles or binaries, further shows variability. These variations do not seem to be exclusively related to binarity and could thus not be systematically interpreted in the framework of a wind-wind collision phenomenon.
\end{abstract}

\section{The early-type star population of NGC\,6231}

\begin{figure}[!t]
\plotfiddle{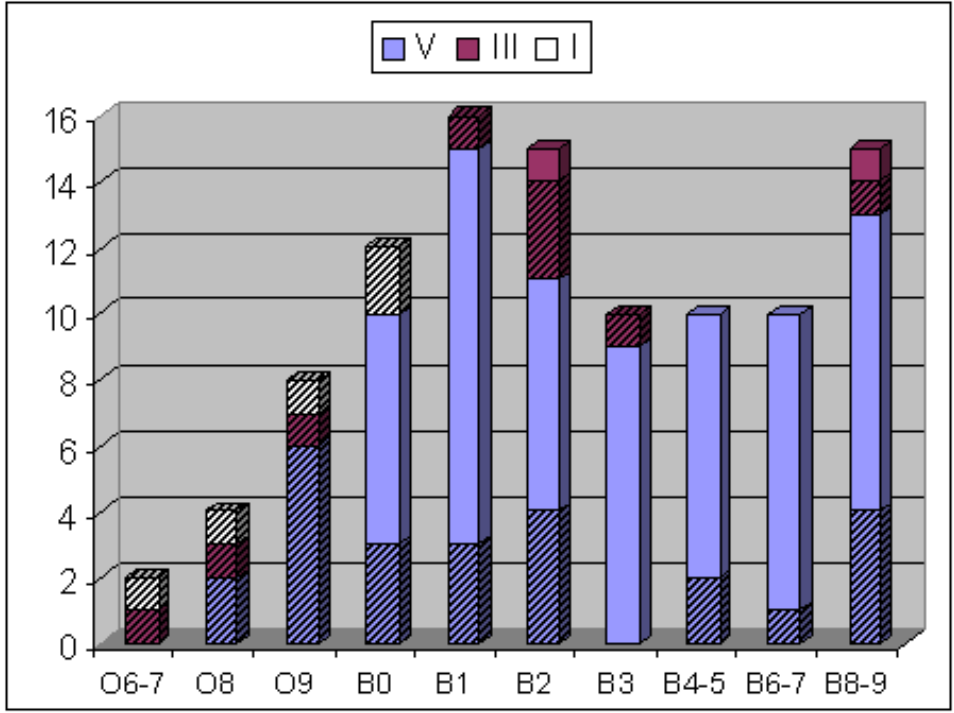}{9cm}{0}{80}{80}{-245}{-185}
\caption{Distribution of the early-type star population within the XMM-Newton f.o.v.\ among the different spectral types and luminosity classes. The shaded regions correspond to detected X-ray emitters.}
\end{figure}

NGC\,6231 is a young  open cluster regarded as the nucleus of the Sco OB1 association \citep{PHC91}. Located at a distance of about 1.7 kpc, it contains a few hundred early-type stars and offers an excellent sample of different spectral types and luminosity classes. Hence NGC\,6231 has been chosen as the target of an XMM-Newton observing campaign in the framework of the guaranteed time of the Optical Monitor consortium. The XMM-Newton field of view (f.o.v.) was centered on the colliding wind binary (CWB) HD\,152248 in the core of the cluster \citep{SRG01, SSG04}. Within the 15 arcmin radius of the f.o.v., 110 early-type stars are reported in the literature, among which 15 O-type objects, either single or multiple, and a Wolf-Rayet system (WR\,79). About one half of the O-type objects are spectroscopic binaries for which we have detected the signature of both components using high resolution high S/N ratio spectra. Among the presumably single O-type stars, another half shows, on time-scales of a few days, small RV shifts of a peak-to-peak amplitude of about 15 to 30 km\,s$^{-1}$. These could not be associated to a possible multiplicity yet and we will therefore consider them as single in the following. Figure 1 presents the distribution of the early-type star population of the cluster among the different spectral types and luminosity classes. 

\section{X-ray observations of NGC\,6231}

The XMM-Newton campaign consisted of six 30-ksec exposures spread over five days, for a total nominal duration  of about 180 ksec. Despite some background flares, this data set  forms one of the deepest X-ray views towards a young open cluster yet performed. Figure 2 presents a combined image of the three EPIC instruments and demonstrates that the field in the X-ray domain is, indeed, very crowded. 

Based on the {\it edetectchain} of the Science Analysis System (SAS) software, we detected 610 X-ray sources in the f.o.v. An optical counterpart within a cross-correlation radius of 3 arcsec could be identified for about 520 of these objects ($\approx85$\%). Most of them seem to belong to the cluster. The estimation of the detection limit is a more tricky issue, the sensitivity of XMM-Newton being clearly not uniform throughout the f.o.v. For NGC\,6231 early-type stars displaying a typical temperature of 0.7\,keV, we estimate this limit to be about 3 to 8 10$^{-15}$\,erg \,s$^{-1}$\,cm$^{-2}$ depending on the position on the detectors. At the distance of the cluster, this yields luminosity limits ranging from 1.0 to 2.5\,10$^{30}$\,erg\,s$^{-1}$. More details on the estimation of the detection limit and on the factors influencing it will be given in \citet{Sana05}.

About 40 X-ray sources are associated with early-type stars and Fig.1 presents their repartition among the early-type population of the cluster. All the O-type stars, both singles and binaries, are detected in the X-ray domain, as are all the supergiants and most of the giants in the f.o.v.  On the contrary, only a small fraction ($\sim$30\%) of the B-type main-sequence stars are detected. We have obtained high resolution snapshot spectra of most of the B-type X-ray emitters and, except in one case, we were not able to detect evidence of binarity. Their RVs are further in good agreement with the mean velocity of the cluster. We can therefore reasonably consider them as singles. About one fourth of the B-type emitters are foreground objects while two other objects are known $\beta$\,Cep stars belonging to the cluster. 

\begin{figure}[!t]
\plottwo{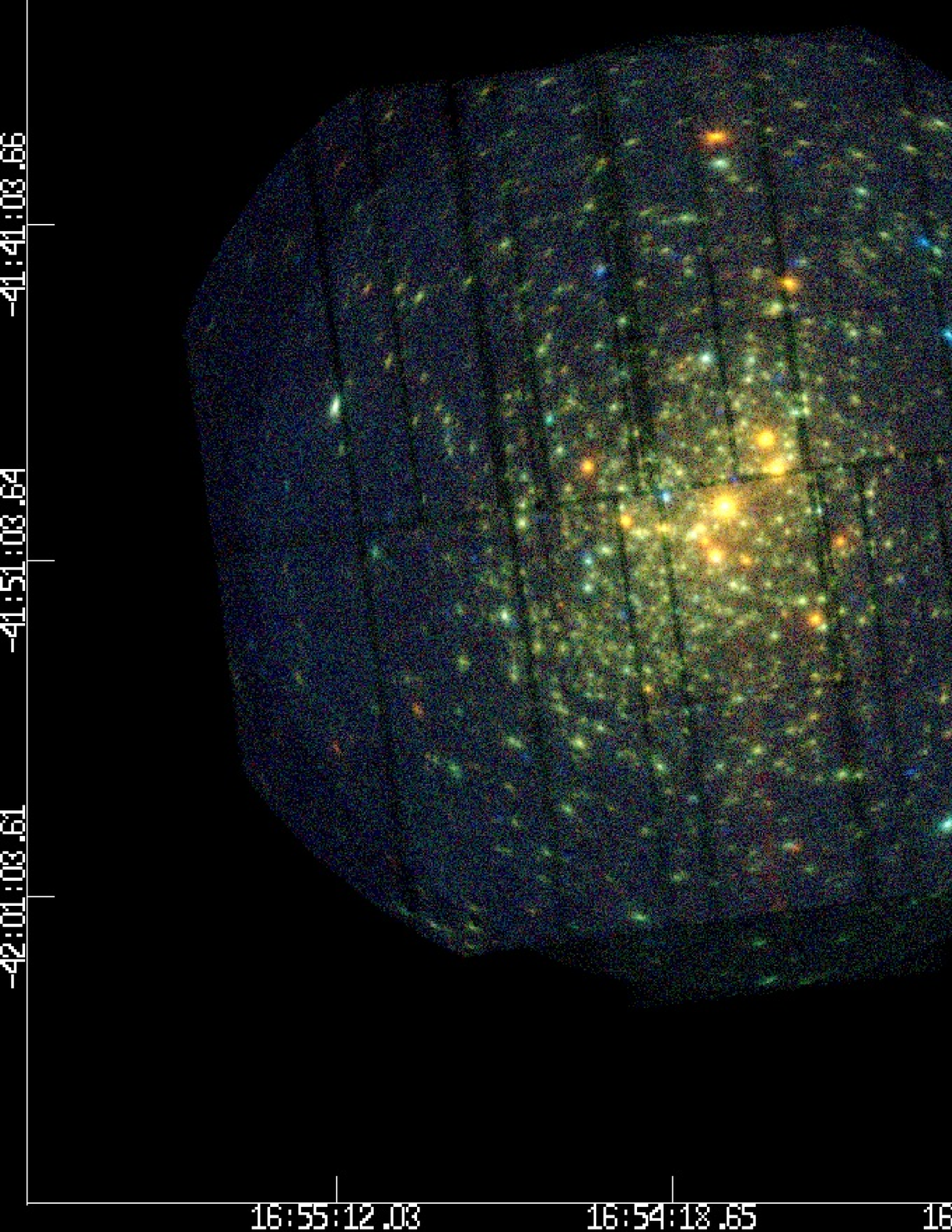}{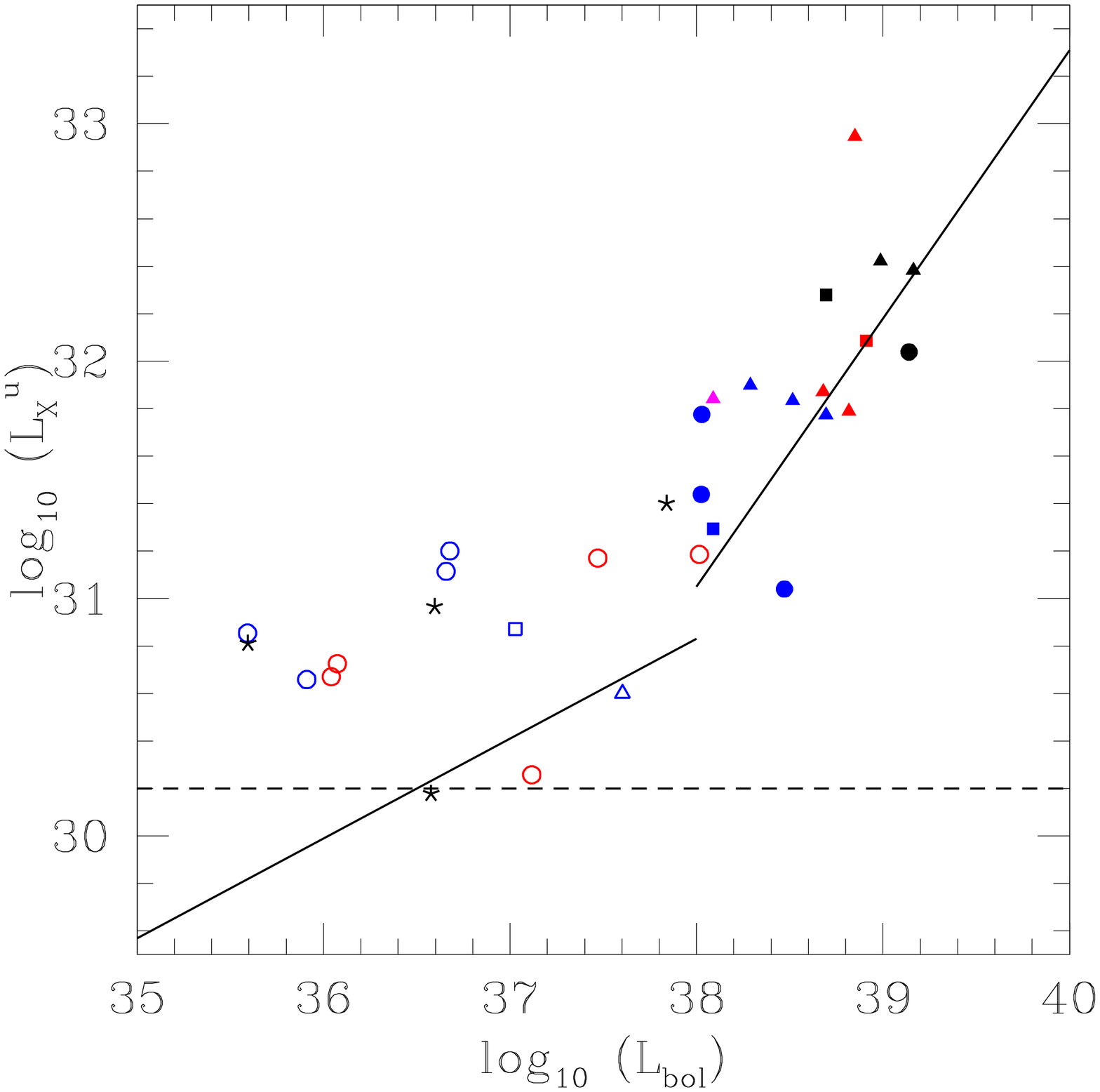}
\caption{{\bf Left:} Combined three-colour X-ray image of NGC\,6231: red [0.5--1.0\,keV]; green [1.0-2.5\,keV]; blue [2.5-10.0\,keV]. This image appears in colour in the electronic version of the paper. {\bf Right:} $L_\mathrm{X}$ vs.\ $L_\mathrm{bol}$ diagram for the early-type stars of NGC\,6231 detected by XMM-Newton. Both axis are expressed in terms of $\log$(erg\,s$^{-1}$). The unabsorbed X-ray luminosities are restricted to the 0.5-10.0\,keV energy band. Solid lines indicate the \citeauthor{BSC97}\ relations for O- and B-type stars in the 0.1-2.0\,keV band. The dashed line gives an estimation of our detection limit for early-type stars. O and B stars are respectively indicated by filled and open symbols. Triangles are for binaries, squares for stars displaying small RV shifts and circles for presumably single stars. The asterisks indicate foreground objects.}
\end{figure}

\section{$L_\mathrm{X} / L_\mathrm{bol}$ relationship}

Among the 40 early-type X-ray emitters, we selected 31 of them that are bright and isolated enough. We extracted their X-ray spectra using the merged event-list of the 6 pointings. We then  estimated their X-ray fluxes by adjusting up to three-temperature optically thin thermal plasma models (\texttt{mekal} models). For this, we adopted individual ISM absorbing columns to reflect the differential reddening across the field \citep*{SBL98}. Most of these X-ray spectra are well described by \texttt{mekal} models with temperatures below 1\,keV. We then computed the unabsorbed X-ray fluxes and, adopting a distance modulus of 11.07 for NGC\,6231, we obtained the X-ray luminosities. Finally, based on the photometry of \citet{SBL98}, we computed the bolometric luminosity of the selected objects.

Figure 2 presents the  $L_\mathrm{X} / L_\mathrm{bol}$ diagram for the early-type stars of NGC\,6231. There is a clear separation between the behaviours of the  O- and  B-stars with a cut-off limit at $L_\mathrm{bol}\approx10^{38}$\,erg\,s$^{-1}$, as previously proposed by \citet {BSC97}. Though the dispersion is quite large, there is no obvious separation between single and binary O-type stars in our sample, except for the CWB HD\,152248 ($\log\left(L_\mathrm{X}\right)\approx33 $). Most of the detected B-stars are far more X-ray luminous than what is expected from the \citeauthor{BSC97}\ relation. However a direct comparison is difficult because of the different energy bands considered. In Fig. 2, the \citeauthor{BSC97}\  relations (obtained in the 0.1-2.0\,keV band) are only given as indications. The non-detection of a large fraction of the B-type star population is more puzzling, indicating that most of them have an X-ray luminosity below about 10$^{30}$\,erg\,s$^{-1}$.  This might be explained by a largely spread B-star emission, in which case we only observed the upper part of the distribution. However, if the dispersion was approximately uniform, we would expect to observe more stars slightly above the detection limit, in the region of the plot between about $\log(L_\mathrm{bol})=36$ and 38. This could suggest a bimodal distribution of the B-type stars in the $L_\mathrm{X} / L_\mathrm{bol}$ diagram. This point clearly deserves more attention in a near future.

\section{X-ray variability}

This last section briefly investigates the variability of the detected X-ray sources. Our XMM-Newton campaign actually consisted of six 30-ksec exposures spread over 5 days. The first aim was to monitor the X-ray emission from the five massive binaries  with a period shorter than 6 days in the f.o.v. These data however provide a good opportunity to probe the variability of various types of X-ray sources on a time-scale of a few days. 

We thus perform a $\chi^2$ variability test. We adopted the null hypothesis of a constant count rate, in the 0.5-10.0\,keV energy band, for each of the six pointings and we required a 90\% confidence level in each EPIC instrument to reject that hypothesis. About 40 \% of all the sources and 33\% of the early-type X-ray emitters present a clear variability, while 25\% do not show any sign of variability in any of the instruments. Requiring a 99\% confidence level does not affect this pattern much since 25\% of the sources are  still reported as variable. This fraction is much larger than previously thought, clearly indicating that variability is a common phenomenon among stellar X-ray emitters. 

Finally, one half of the short period O-type binaries present definite variability while another fourth display some sign of variability in at least one instrument. The emission from the WR system remains approximately constant throughout the six pointings, but we actually covered only 5 days out of its 9-day period. O-star X-ray variability does not seem to be exclusively related to binarity and could therefore not be uniquely explained by a wind-wind interaction scenario.

\section{Conclusions}

Using the XMM-Newton X-ray observatory, we performed one of the deepest X-ray investigations towards a young open cluster. Of an effective duration of at least 170\,ksec, our campaign reveals a very crowded field with 610 detected sources, among which 42 early-type stars.
We obtained the $L_\mathrm{X} / L_\mathrm{bol}$ relation for the early-type emitters of the cluster. We confirm the clear dichotomy between the behaviours of the  O- and B-type stars. No systematic X-ray overluminosity could be assigned to early-type binaries. Most of the B-type stars of the cluster are  more X-ray luminous than expected from the \citeauthor{BSC97}\ relation. 
We finally investigated the variability of the X-ray sources and showed that variability affects a large fraction of the detected sources. 

\acknowledgments{
This work was based on observations collected with XMM-Newton, an ESA Science Mission with instruments and contributions directly funded by ESA Member States and the USA (NASA). HS is a Research Fellow FNRS (Belgium); EG and GR are Research Associates FNRS (Belgium). HS is grateful to the FNRS (Belgium), to the ``Patrimoine de l'Universit\'e de Li\`ege'' and to the meeting SOC for contributing to his participation at this workshop. This research is also supported in part by contract P5/36 ``P\^ole d'Attraction Interuniversitaire'' (Belgian Federal Science Policy Office) and through the PRODEX XMM-OM and INTEGRAL projects. The SIMBAD database has been consulted for the bibliography.  }

\end{document}